\documentclass[superscriptaddress,reprint,aps,prb]{revtex4-1}

\usepackage{amssymb, amsmath}
\usepackage{graphicx}
\usepackage{multirow}

\usepackage{color}

\begin{document}

\preprint{APS/PRB}

\title{Strain-induced magnetic phase transition in SrCoO$_{3-\delta}$ thin films}

\author{S.~J. Callori}
\email{s.callori@unsw.edu.au}
\affiliation{School of Physics, The University of New South Wales, Sydney, NSW 2052, Australia}
\affiliation{The Bragg Institute, Australian Nuclear Science and Technology Organization, Lucas Heights, NSW 2234, Australia}
\author{S. Hu}
\affiliation{School of Materials Science and Engineering, The University of New South Wales, Sydney, NSW 2052, Australia}
\author{J. Bertinshaw}
\affiliation{School of Physics, The University of New South Wales, Sydney, NSW 2052, Australia}
\author{Z. Yue}
 \affiliation{Australian Institute for Innovative Materials and Institute for Superconducting and Electronic Materials,
University of Wollongong, North Wollongong, NSW 2522, Australia}
\author{S. Danilkin}
\affiliation{The Bragg Institute, Australian Nuclear Science and Technology Organization, Lucas Heights, NSW 2234, Australia}
\author{X.~L. Wang}
 \affiliation{Institute for Superconducting and Electronic Materials, University of Wollongong, North Wollongong, 
 New South Wales 2500, Australia}
\author{V. Nagarajan}
\affiliation{School of Materials Science and Engineering, The University of New South Wales, Sydney, NSW 2052, Australia}
\author{F. Klose}
\affiliation{The Bragg Institute, Australian Nuclear Science and Technology Organization, Lucas Heights, NSW 2234, Australia}
\affiliation{Department of Physics and Materials Science, City University of Hong Kong, Hong Kong SAR, China}
\author{J. Seidel}
\affiliation{School of Materials Science and Engineering, The University of New South Wales, Sydney, NSW 2052, Australia}
\author{C. Ulrich}
\email{c.ulrich@unsw.edu.au}
\affiliation{School of Physics, The University of New South Wales, Sydney, NSW 2052, Australia}
\affiliation{The Bragg Institute, Australian Nuclear Science and Technology Organization, Lucas Heights, NSW 2234, Australia}

\date{\today}

\begin{abstract}
It has been well established that both in bulk at ambient pressure and for films under modest strains, cubic
SrCoO$_{3-\delta}$  ($\delta < 0.2$) is a ferromagnetic metal. Recent theoretical work, however, indicates that
a magnetic phase transition to an antiferromagnetic structure could occur under large strain accompanied by a
metal-insulator transition. We have observed a strain-induced ferromagnetic to antiferromagnetic phase transition
in SrCoO$_{3-\delta}$ films grown on DyScO$_3$ substrates, which provide a large tensile epitaxial strain, as
compared to ferromagnetic films under lower tensile strain on SrTiO$_3$ substrates. Magnetometry results demonstrate
the existence of antiferromagnetic spin correlations and neutron diffraction experiments provide a direct evidence
for a G-type antiferromagnetic structure with Ne\'el temperatures between $T_N \sim 135\,\pm\,10\,K$ and
$\sim 325\,\pm\,10\,K$ depending on the oxygen content of the samples. Therefore, our data experimentally confirm
the predicted strain-induced magnetic phase transition to an antiferromagnetic state for SrCoO$_{3-\delta}$ thin films
under large epitaxial strain.
\end{abstract}


\maketitle

The broad range of transition metal oxide functionalities, including superconductivity, magnetism, and ferroelectricity,
can be tuned by the careful choice of parameters such as strain, oxygen content, or applied electric and magnetic fields
\cite{Jeen13_1,Jeen13_2,Yang09,Seidel11,Takeda72,Choi13,Jeen13_3,Ikeda14,Scullin10}.
This tunability makes transition metal oxide materials ideal candidates for use in developing novel information and
energy technologies \cite{Ramesh07,Martin10}. SrCoO$_3$ is a particularly interesting system for investigation.
SrCoO$_{3-\delta}$ has long been studied due to its propensity to form oxygen-vacancy-ordered structures as the oxygen
content is decreased. The system undergoes well-defined structural phase transitions between distinct topotactic phases, from
a cubic perovskite phase at SrCoO$_{3}$ to brownmillerite SrCoO$_{2.5}$. The ties between the structural and functional
properties of the material are obvious as a magnetic phase transition from ferromagnetic (FM) SrCoO$_{3.0}$ with
T$_C$ = 280-305\,K  to antiferromagnetic (AFM) SrCoO$_{2.5}$ with T$_N$ = 570\,K accompanies the structural transition
\cite{Takeda72,Taguchi79,Bezdicka93,Long11}. This is similar to the case of SrFeO$_{3-\delta}$, which has also been
demonstrated to undergo oxygen vacancy ordering with magnetic phase transitions related to the structure and Fe charge
ordering \cite{Hodges00,Lebon04,Reehuis12}.

In addition to oxygen stoichiometry, other possibilities, such as strain or applied magnetic or electric fields, may be used
to tune the system. Lee and Rabe have simulated the effect of epitaxial strain on SrCoO$_{3.0}$ and predict a large polar
instability resulting in a dependence of the magnetic structure on strain \cite{Lee11_1,Lee11_2}.
Their results show that the magnetic state can be controlled by the amount of compressive or tensile strain applied.
An AFM-FM transition is predicted at both, tensile strain of $\sim$2.0\,\% and compressive strain of approximately -0.8\,\%,
which is caused through the Goodenough-Kanamori rules as a consequence of simultaneous structural phase transitions
between phases with different distortions and rotational patterns of the CoO$_6$ octahedra. \cite{Lee11_1}
Furthermore, the magnetic phase transitions would be accompanied by ferroelectric (FE) and metal-insulator transitions due to
the strong coupling between the lattice, electric polarization, and electronic band gap \cite{Lee11_1,Lee11_2} and could
potentially enable SrCoO$_3$ to be incorporated into multiferroic heterostructures \cite{Song14}. If these predictions hold,
thin film epitaxy can be used to engineer SrCoO$_{3}$ films near a calculated strain phase boundary, where an external magnetic
or electric field could be used to drive the system from an insulating AFM/FE to a FM/metallic phase. This method of strain
engineering highlights the potential of developing SrCoO$_3$ as a novel multiferroic material and demonstrates the coupling
of three controllable properties: magnetism, ferroelectricity, and strain.

In this Rapid Communication we demonstrate a strain-induced FM-AFM phase transition in SrCoO$_{3-\delta}$ (SCO),
with $\delta < 0.2$, as is predicted theoretically. Due to lattice mismatch, by growing SCO films on DyScO$_3$
(DSO) substrates, we were able to apply a high degree of in-plane tensile strain. SCO films were also prepared
on SrTiO$_3$ (STO) substrates in order to compare the SCO/DSO sample with that under a lower in-plane tensile
strain. The aim, then, is to determine the magnetic ground state of the two samples with different in-plane
strains using neutron diffraction and magnetometry.

Previous results have shown that the magnetic phase transition temperatures are highly dependent on the strain state and
oxygen deficiency of SCO. In its bulk form stoichiometric SrCoO$_{3}$ is a ferromagnetic metal. The spins of the Co$^{4+}$ ion
are arranged in the intermediate spin state forming a S = 3/2 system and a magnetic moment of up to 2.5\,$\mu_B$ has been
observed. \cite{Long11} The ferromagnetic ground state is a consequence either directly of the double exchange interaction
or caused by a hybridization with the oxygen atoms due to the negative charge transfer gap, forming a 3d$^6$-L ground state
with FM properties. \cite{Potze95}. The observed ordering temperature of FM SrCoO$_{3.0}$ ranges from $\sim$\,280\,-\,305\,K
\cite{Bezdicka93,Long11} for the bulk, to even lower temperatures for films under tensile strain. As the degree of strain is
increased, the FM transition temperature decreases, from $\sim$\,250\,K under low tensile strains on
(LaAlO$_3$)$_{0.3}-$(SrAl$_{0.5}$Ta$_{0.5}$O$_3$)$_{0.7}$ (0.9\,\% strain) \cite{Jeen13_2}, to $\sim$\,200\,K for thin FM
films of SCO under a more intermediate tensile strain on STO substrates (1.8\,\% strain)\cite{Jeen13_1}. Additionally, in SCO
the oxygen vacancies order systematically, which results in the development of various distinct topotactic phases. Each phase
possesses a different magnetic ground state, i.e. FM with T$_C$ = 220\,K for SrCoO$_{2.875}$, FM with T$_C$ = 160\,K for
SrCoO$_{2.75}$, and AFM with T$_N$ = 570\,K for the brownmillerite phase SrCoO$_{2.5}$.

\begin{figure}[t!]
\begin{center}
  \includegraphics[width=6.9cm]{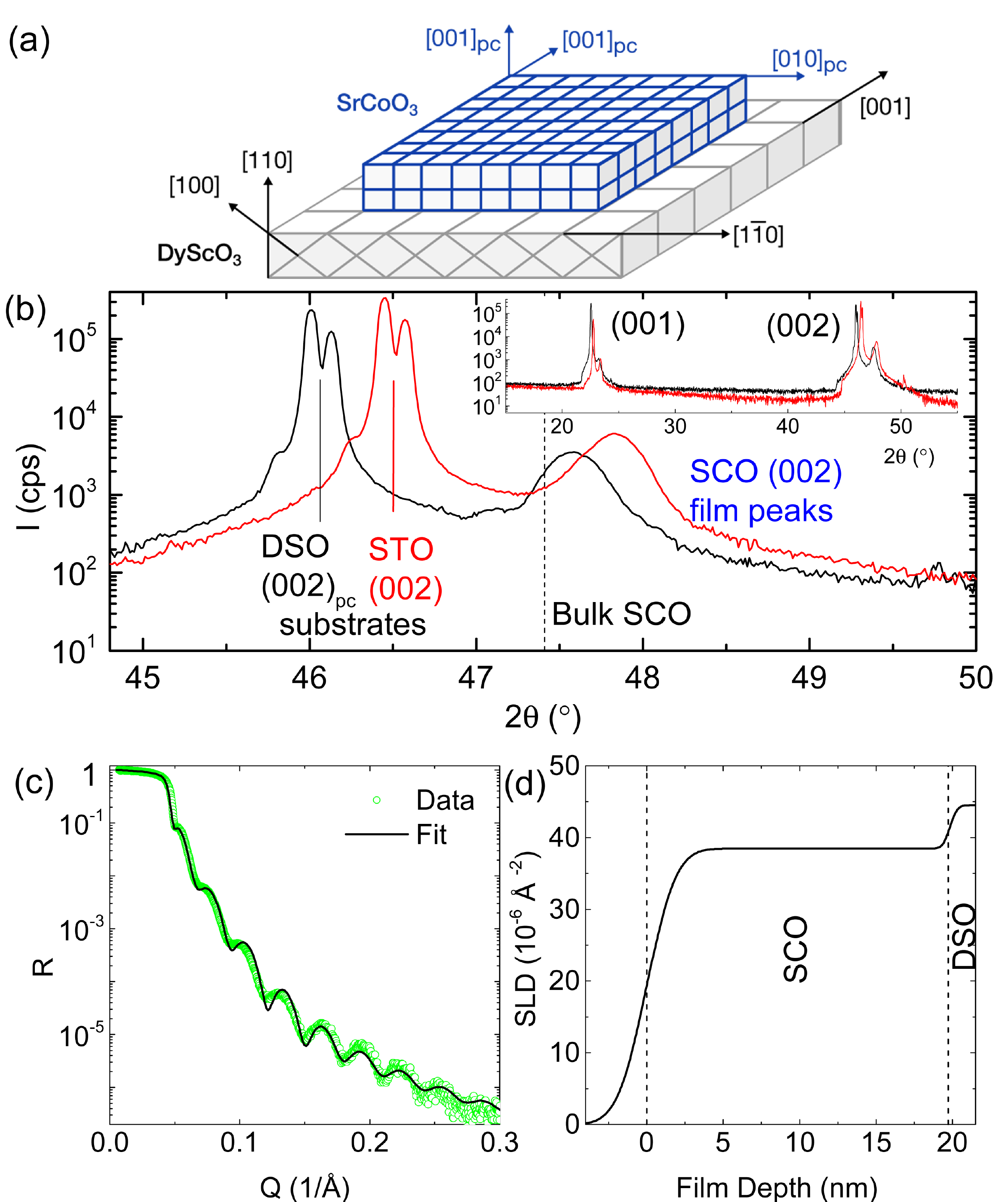}
  \caption{\textit{(color online)
  (a) Schematic of the lattice directions of the SCO film and DSO substrate. (b) x-ray diffraction of
  SCO/DSO (black) and SCO/STO (red) showing the (002)$_{pc}$ substrate peaks and (002)$_{pc}$ film peaks.
  The dashed line denotes the location of the bulk (002)$_{pc}$ SCO peak.
  The inset presents the entire scan along the (pseudo)cubic (00l)$_{pc}$ directions of the substrates.
  (c) x-ray reflectivity of the 20 nm thick SCO/DSO film. (d) The scattering length density (SLD) across
  the SCO/DSO film (resulting from the analysis of (c)).}}\label{Fig1}
  \end{center}
\end{figure}

A wide range of both tensile and compressive strain is achievable in SCO films through prudent choice of substrate,
with similarly structured perovskite oxides providing the capability for epitaxial growth. At room temperature,
SrCoO$_{3}$ is cubic with $a = b = c = 3.835$\,\AA  \cite{Bezdicka93} and DSO is orthorhombic with $a =5.4494$\,\AA,
$b = 5.7263$\,\AA, and $c =\,7.9170$ \AA. \cite{Lifer04} When considering DSO as a pseudocubic with $a_{pc} = 3.952 $\,\AA  \
(note that the pseudocubic \emph{pc} notation will be used throughout this paper), the rigid substrate leads to a large
in-plane tensile strain of 2.7\,\% on a fully strained, lattice-matched SCO film. To compare the SCO/DSO sample to that under
a lower in-plane tensile strain of 1.8\,\%, films were also deposited on STO (cubic with $a = b = c = 3.905$\,\AA \ \cite{Lytle64}).
A schematic of the film and substrate lattice directions when grown on DSO is shown in Fig. \ref{Fig1} (a), where the SCO cubic
lattice directions are the same as the pseudocubic DSO directions.

SCO films were grown epitaxially on DSO (110)-terminated ((001)$_{pc}$) and STO (100)-terminated substrates using a
pulsed laser deposition (PLD) system with a $248\,\mathrm{nm}$ wavelength KrF excimer laser with conditions described
in Ref. \onlinecite{Hu14}. On DSO, the prepared films were nominally 20\,nm thick, and those deposited on STO were 40\,nm
thick. The thicknesses were chosen to maximize the film thickness while still remaining lattice-matched to each substrate.
The films were initially characterized by laboratory x-ray diffraction and reflectivity using Cu-K$_\alpha$ radiation
(instrument X-Pert MRD Pro), as is shown in Fig. \ref{Fig1} (b) and (c). For films grown on both DSO and STO, the locations
of the (001) and (002) SCO film Bragg peaks occur at higher $2\theta$ values than the bulk (dashed line in Fig. \ref{Fig1} (b)),
indicating smaller \emph{c} lattice parameters and thus confirming that the films are under in-plane tensile strain. The
larger $2\theta$ position for the SCO/DSO sample indicates a smaller out-of-plane lattice parameter than SCO/STO, which is
expected due to the higher tensile strain on DSO. Our previous results indicate that the films are coherently epitaxially
strained to the substrate, with the same in-plane lattice parameters and the films are therefore under a large in-plane
tensile strain. The corresponding x-ray reciprocal space maps are presented in Ref. \onlinecite{Hu14}. It should be
noted that any potential oxygen vacancies in SCO can also affect the out-of-plane lattice parameter. However, due to
the lack of any superlattice reflections in the collected x-ray diffraction pattern (inset of Fig. \ref{Fig1}(b)) and
x-ray reflectivity scans (Fig. \ref{Fig1} (c)) no oxygen vacancy super structures could be identified,
which indicate that the films are near a $\delta\,\approx\,0$ oxygen stoichiometry. \cite{Ikeda14, Nemudry96, Hu14}.
Additional x-ray absorption spectroscopy measurements did indicate that the oxygen content is close to optimal doping
with $\delta \leqslant 0.18$ \cite{Hu14}.

To investigate the nature of the magnetic structure of the SCO films, magnetometry and neutron diffraction experiments were
performed. For elastic neutron scattering the triple-axis spectrometer TAIPAN at the Bragg Institute, ANSTO, Sydney,
Australia was used. This instrument was selected for its excellent signal-to-noise ratio and low background, which are necessary for
measuring the magnetization signal arising from nanometre sized thin films by neutron scattering. The instrument was set up
for standard elastic diffraction with a vertically focused highly-ordered pyrolytic-graphite monochromator, with a fixed incident
and final wavelength of $\lambda = 2.347\,\mathrm{\AA}$. In order to minimize the resolution and background,
40'' collimation was used. Two pyrolytic graphite filters (each 2'' thick) were placed before and after the sample in
the neutron path in order to suppress contamination from second order scattering.

\begin{figure}[t!]
\begin{center}
  \includegraphics[width=6.9cm]{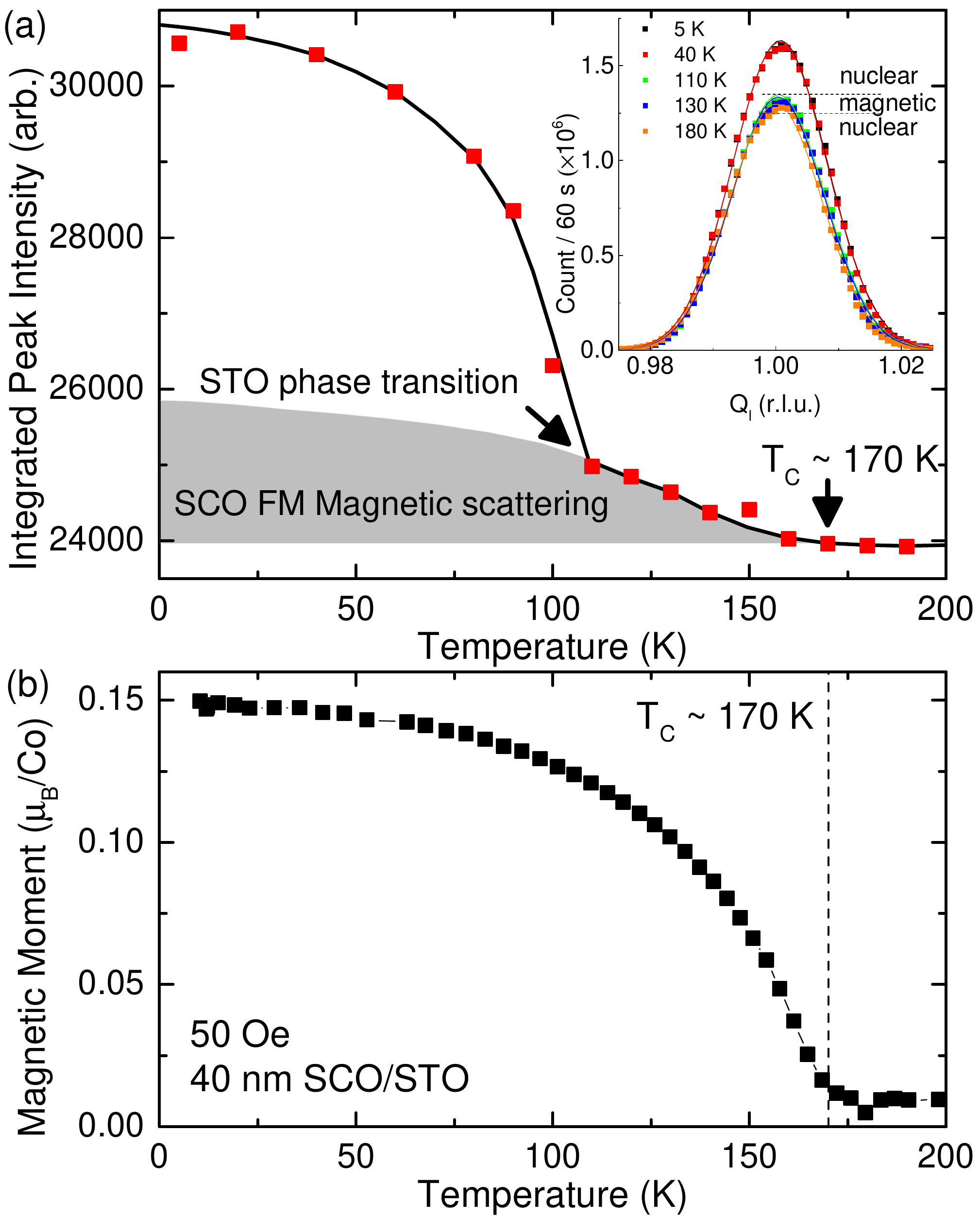}
  \caption{\textit{(color online)
  (a) Temperature dependence of the neutron diffraction intensity of the (001)$_{pc}$ Bragg peak for
  the 400 nm thick SCO/STO sample. The the black line serves as a guide to the eye
  (inset: raw data with Gaussian fits for selected temperatures).
  (b) In-plane measurement of the magnetic moment as a function of temperature (black squares) and integrated
  intensity of the (001) neutron diffraction scans (red circles).}}\label{Fig2}
  \end{center}
\end{figure}

Previous results on epitaxial SCO grown on STO show FM behavior \cite{Jeen13_1,Jeen13_2} which is in agreement with
the predictions of Lee and Rabe.\cite{Lee11_1,Lee11_1} Utilizing the interaction of neutrons with both atomic nuclei
and electron spins, elastic neutron diffraction scans were performed on a 40 nm thick SCO film on STO substrate in order
to confirm the FM ground state on the film. Temperature dependence of the integrated neutron diffraction intensity obtained
from $Q_l$ scans of the (001) Bragg peak are shown in Fig. \ref{Fig2}(a). Due to the lower degree of strain on SCO and,
as compared to the x-ray diffractometer, lower instrumental resolution of the neutron diffractometer, this scan effectively
includes both the substrate and film Bragg peaks. As the STO substrate lacks a magnetic transition, the increase in intensity
of the peak with temperatures below $\sim$\,170\,K can be attributed to the development of FM order in the system. This
agrees well with the reported transition temperatures for SCO films on STO ($\sim$\,200 K) \cite{Jeen13_1,Jeen13_2}.
Due to the cubic-to-tetragonal structural phase transition of the STO substrate, below 110\,K \cite{Shirane69} there is
an additional contribution to the intensity of the (001) Bragg reflection arising from the STO.

Magnetometry measurements were performed on a Quantum Design 14 T PPMS and Quantum Design MPMS (SQUID) on the SCO/STO film
which confirm the FM order in SCO below 170\,K. The resulting temperature dependence of the in-plane magnetization is shown
in Fig. \ref{Fig2} (b). Note that the data were taken at an applied field of 50 Oe, where the magnetic moment of SCO is not
saturated, but larger Co moments of 1.1\,$\mu_B/f.u.$ were measured under a 5000\,Oe field. In bulk SrCoO$_{3.0}$ a total
moment of 2.5\,$\mu_B/f.u.$ was observed \cite{Long11}. Both the magnetization and the neutron diffraction results for the
SCO/STO system confirm that under a moderate in-plane tensile strain the SCO film is indeed FM with a transition temperature
of roughly 170 K, in agreement with previous results \cite{Jeen13_1}.

\begin{figure}[t!]
\begin{center}
  \includegraphics[width=6.9cm]{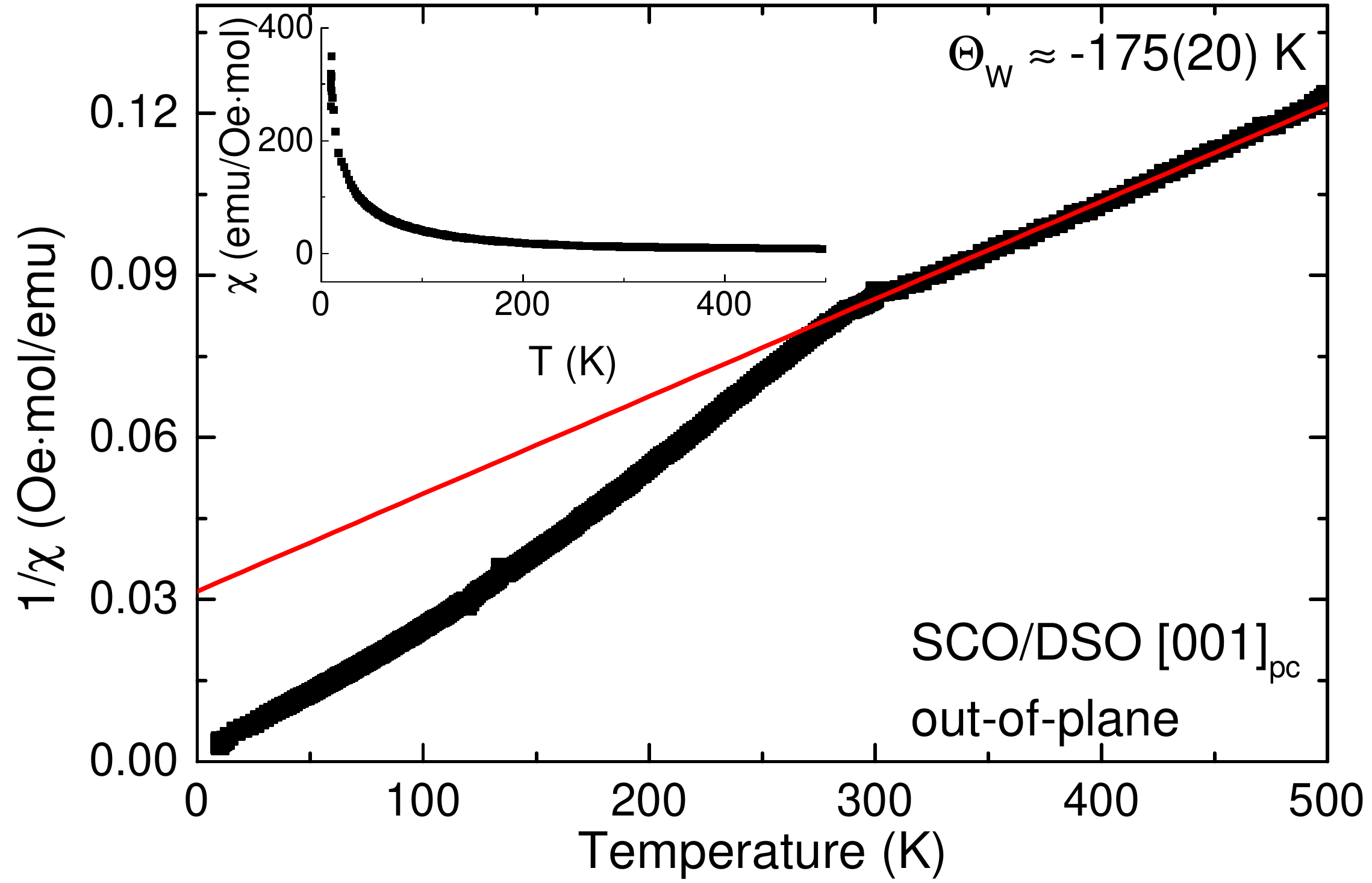}
  \caption{(color online)
  \textit{$1/\chi$ measured for the [110] ([001]$_{pc}$) out-of-plane direction of the 20\,nm thick SCO/DSO film.
  The red line shows the Curie-Weiss fit to the high temperature data. The inset shows the corresponding
  susceptibility, $\chi$.}}\label{Fig3}
  \end{center}
\end{figure}

Magnetization measurements were performed to search for the existence of AFM spin correlations in SCO grown on DSO substrate.
$1/\chi$ as a function of temperature for SCO/DSO is plotted in Fig. \ref{Fig3}. The DSO paramagnetic signal was accounted
for by subtracting the magnetization signal measured on a pure substrate. At low temperatures ($\lesssim$\,30\,K), there is
a large contribution from the DSO substrate caused by spin fluctuations near the DSO magnetic phase transition, occurring at
3\,K.\cite{Ke09}

The $1/\chi$ high temperature data (above 300\,K) were fit to a linear Curie-Weiss law, $1/\chi = \frac{T-\Theta_W}{C}$,
where $\Theta_W$ is the Weiss temperature (see red line in Fig. \ref{Fig3}). A $\Theta_W$ value of $-175\,\pm\,20$\,K is
found. The negative $\Theta_W$ value is a first indication of AFM spin correlations in the SCO/DSO film and therefore an AFM
state would be expected to occur around +175\,K. Also observable is a deviation from the fit below $\sim$\,280\,K, which
may indicate the existence of a minor FM impurity phase within the SCO film. Additionally, due to the strain on the film,
an anisotropy in the magnetic properties would most likely be expected as a consequence of the different orbital ordering
pattern in-plane and out-of-plane. \cite{Konishi99,Aruta06,Nath99,Xiong05}

\begin{figure}[t!]
\begin{center}
  \includegraphics[width=6.9cm]{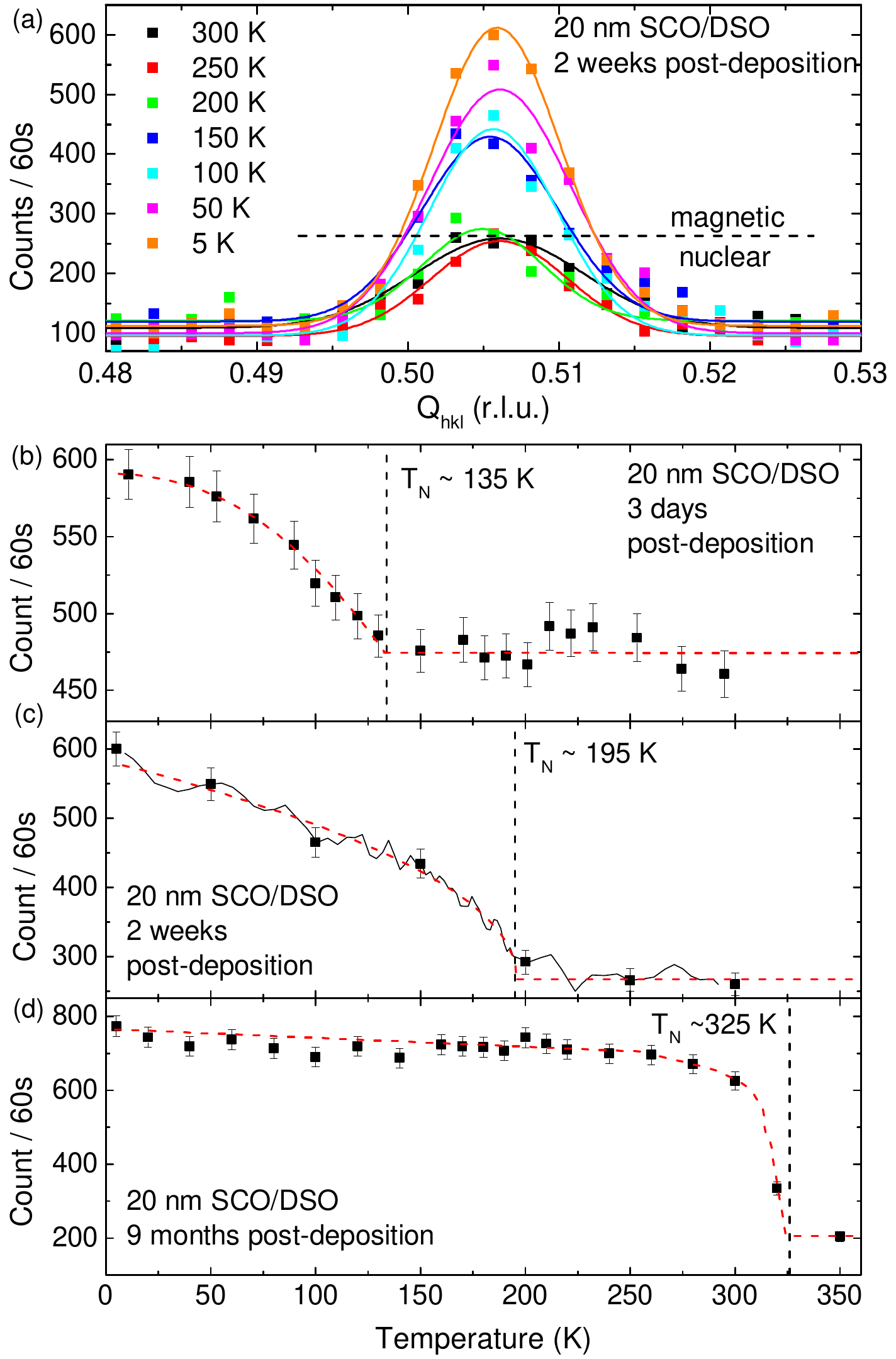}
	\caption{(color online)
    \textit{(a) Neutron diffraction scans of the $(\frac{1}{2} \frac{1}{2} \frac{1}{2})_{pc}$ SCO Bragg peak
    of the 20\,nm thick SCO film grown on DSO. Intensity of the $(\frac{1}{2} \frac{1}{2} \frac{1}{2})_{pc}$
    reflection of the SCO/DSO film (b) 3 days, (c) 2 weeks, and (d) 9 months post-deposition.
    The black squares are peak intensities of the scans derived from Gaussian fits to the data as indicated in (a).
    The solid line in (c) corresponds to the intensity of the peak maximum recorded as the temperature was raised from
    5\,K to room temperature. The dashed red lines serve as guides to the eye.}}
    \label{Fig4}
  \end{center}
\end{figure}

Since the Curie-Weiss fit to the magnetometry data is only capable of providing information about the short range
interactions between moments in the paramagnetic phase, neutron diffraction experiments were performed in order to
solidly determine the ordered magnetic structure of the strained SCO/DSO film. The substrate was aligned along the
$[110]_{pc} \times [001]_{pc}$ scattering plane allowing access to all Q$_{hhl}$ reflections. Due to the prediction
that the film would have a G-type AFM structure \cite{Lee11_1}, the $(\frac{1}{2} \frac{1}{2} \frac{1}{2})_{pc}$
peak region was monitored as a function of temperature (see Fig. \ref{Fig4}).

Fig. \ref{Fig4} (a) shows $Q_{hkl}$ neutron scattering scans of the G-type AFM Bragg peak taken over a range of
temperatures of the 20\,nm thick SCO/DSO film two weeks post-deposition. Figures \ref{Fig4} (b),(c), and (d) show
the intensity of the $(\frac{1}{2} \frac{1}{2} \frac{1}{2})_{pc}$ Bragg reflection for the thin film samples
three days, two weeks, and nine months  after deposition, respectively. In each case the intensity of the
$(\frac{1}{2} \frac{1}{2} \frac{1}{2})_{pc}$ Bragg reflection increases at low temperatures, indicating the onset of
G-type AFM order. The intensity above the transition originates from a weakly allowed nuclear Bragg peak of the DSO
substrate and a formerly forbidden Bragg reflection of the film which becomes active due to the epitaxial strain causing
an orthorhombic distortion. The magnetic phase transition temperatures are $135\,\pm\,10$\,K, $195\,\pm\,10$\,K, and
$325\,\pm\,10$\,K for the samples measured three days, two weeks, and nine month after deposition, respectively.
In particular the transition temperature of the second sample agrees well with the $\Theta_W$ value obtained from
magnetometry, where the sample was measured one week after growth. The magnetic phase transition temperature
increases systematically with the age of the sample, indicating that the oxygen content has changes with time.
Additional x-ray diffraction measurements on the aged samples did not show any superstructure reflections, demonstrating
that the films remained in the perovskite phase. However, a relaxation of the c-axis lattice parameter was observed.

A rough estimate based on the comparison of the intensities of the magnetic Bragg peak of the FM SCO film grown on STO with
the AFM SCO on DSO, assuming that the spins are in-plane, yields a maximum magnetic moment of 0.5\,$\mu_B$ per Co-ion
for the SCO/DSO sample two week post-deposition. Note that the determination of the precise magnet moment would
require the scaling to a nuclear Bragg reflection. However, this was not possible since the nuclear Bragg reflections of SCO
fall on top of the dominating Bragg reflections of the DSO substrate. The reduced magnetic moment probably indicates that
the AFM order of the SCO film on DSO arises from the low-spin configuration of the Co$^{4+}$ 3$d^5$ electrons in the Mott
insulating regime. This is similar to LaCoO$_3$, which undergoes a transition from a low-spin to a mixed-spin state with
increasing temperature \cite{Have06}. The magnetic contribution to the $(\frac{1}{2} \frac{1}{2} \frac{1}{2})_{pc}$
Bragg reflection increases systematically with the age of the film, indicating an increase of the magnetic moment as a consequence
of the changed oxygen content. Additional scans of the (001)$_{pc}$ peak ruled out the existence of further FM contributions
within the experimental resolution (data not shown here). Over the temperature range of the experiment, the DSO substrate is
paramagnetic with no structural phase transitions \cite{Ke09}. Therefore, no contribution from the substrate is expected in
these measurements.

The noticeable increase in intensity below $T_N$ evidently is due to the AFM signal of the SCO film and demonstrates a
phase transition from paramagnetism to G-type AFM. This behavior has not previously been reported for SCO films under
smaller tensile strains and does not correspond to any other known magnetic phases of bulk SCO. The only other identified
AFM phase is brownmillerire SrCoO$_{2.5}$, with T$_N$ = 570\,K, but XRD and XRR rule out oxygen vacancy superstructure
reflections in the SCO/DSO sample (see Fig. \ref{Fig1}), which would indicate the presence of this phase.

In conclusion, control over the magnetic phase of SCO thin films was realized through the application of a large
epitaxial strain, via substrate choice. As a result these films displayed i.) a well known FM behavior under
low tensile strain when grown on STO, which has been well established for SCO in bulk \cite{Taguchi79,Bezdicka93,Long11}
(T$_C$ = 280-305\,K) and in epitaxial films under a lower degree of tensile strain \cite{Jeen13_1,Jeen13_2,Rueckert13}
and ii.) G-type AFM ordering under large tensile strain when deposited on DSO. While the transition temperature of
the SCO/DSO film is close to that of the SCO/STO sample, there is a clear distinction between the magnetic states of
antiferromagnetic SCO/DSO and ferromagnetic SCO/STO.

The results presented here represent the first step in obtaining new functionalities in SCO. These observations confirm
the predicted strong coupling between structure (via strain) and magnetism in SCO and demonstrate how this coupling can
be used to tailor the magnetic properties of the material. This result also serves as the first confirmation of the model
of Lee and Rabe \cite{Lee11_1,Lee11_2}, which opens the possibility of investigating SCO for further coupling between
magnetism, lattice structure, and electric polarization and the ultimate development of SCO as a novel multiferroic,
combining control of spin order, charge degrees of freedom via metal-insulator transition, and electrical polarization
through the application of epitaxial strain.

\begin{acknowledgments}
We acknowledge support by the Australian Research Council (ARC) under grant numbers FT110100523, DP110105346, DP140100463,
and DP140102849. This work was also supported by the Australian Institute of Nuclear Science and Engineering (AINSE) and
the National Research Foundation of Korea funded by the Korean Government (NRF-2013S1A2A2035418).
\end{acknowledgments}

\end{document}